\documentclass[iop]{emulateapj}
\usepackage{epstopdf}
\usepackage{graphicx}
\usepackage{url}
\usepackage{natbib}
\usepackage{amsmath}
\usepackage{comment}
\providecommand{\mu}{$\Delta m_{15}(U)$}

\begin{document}
%%%%%%%%%% Title & Author %%%%%%%%%%%%%%%%%%%%%%%%%%%%%%%%%%%%%%%%%%%%%%%%%%%%%%%
\title{ASASSN-15\lowercase{lh}: A Superluminous Ultraviolet Rebrightening \\
Observed by Swift and Hubble*}\thanks{*Based on observations with the NASA/ESA Hubble Space Telescope obtained at the
Space Telescope Science Institute, which is operated by the Association of Universities
for Research in Astronomy, Incorporated, under NASA contract NAS5-26555.}

\author{Peter~J.~Brown\altaffilmark{1}, Yi Yang\altaffilmark{1}, Jeff Cooke\altaffilmark{2}, Melanie Olaes\altaffilmark{3}, Robert M. Quimby\altaffilmark{3,4}, \\
Dietrich Baade\altaffilmark{5},  Neil Gehrels\altaffilmark{6}, Peter Hoeflich\altaffilmark{7}, Justyn Maund\altaffilmark{8},  \\
Jeremy Mould\altaffilmark{2}, Lifan Wang\altaffilmark{1}, \& J. Craig Wheeler\altaffilmark{9}}

\altaffiltext{1}{George P. and Cynthia Woods Mitchell Institute for Fundamental Physics \& Astronomy, 
Texas A. \& M. University, Department of Physics and Astronomy, 
4242 TAMU, College Station, TX 77843, USA }            
\altaffiltext{2}{Centre for Astrophysics \& Supercomputing, Swinburne University, Hawthorn VIC 3122, Australia}

\altaffiltext{3}{Department of Astronomy, San Diego State University, San Diego, CA 92182, USA}

\altaffiltext{4}{Kavli IPMU (WPI), UTIAS, The University of Tokyo, Kashiwa, Chiba 277-8583, Japan}

\altaffiltext{5}{European Organisation for Astronomical Research in the Southern Hemisphere (ESO), 
Karl-Schwarzschild-Str. 2,
85748 Garching b. M\"unchen, Germany}

\altaffiltext{6}{NASA Goddard Space Flight Center, Greenbelt, MD 20771, USA}

\altaffiltext{7}{Department of Physics, Florida State University, Tallahassee, FL 32306, USA}

\altaffiltext{8}{Department of Physics and Astronomy
F39 Hicks Building, Hounsfield Road
Sheffield, S3 7RH, United Kingdom}

\altaffiltext{8}{Department of Astronomy, University of Texas at Austin, Austin, TX 78712, USA}

\setcounter{footnote}{0}

%%%%%%%%%% Abstract %%%%%%%%%%%%%%%%%%%%%%%%%%%%%%%%%%%%%%%%%%%%%%%%%%%%%%%%%%%%
\begin{abstract}

We present and discuss ultraviolet and optical photometry from the Ultraviolet/Optical Telescope and X-ray limits from the X-Ray Telescope on Swift and imaging polarimetry and ultraviolet/optical spectroscopy with the Hubble Space Telescope of ASASSN-15lh. It has been classified as a hydrogen-poor superluminous supernova (SLSN I) more luminous than any other supernova observed.  ASASSN-15lh is not detected in the X-rays in individual or coadded observations.  From the polarimetry we determine that the explosion was only mildly asymmetric.  We find the flux of ASASSN-15lh to increase strongly into the ultraviolet, with a ultraviolet luminosity a hundred times greater than the hydrogen-rich, ultraviolet-bright SLSN II SN~2008es.   We find objects as bright as ASASSN-15lh are easily detectable beyond redshifts of $\sim$4  with the single-visit depths planned for the Large Synoptic Survey Telescope.   Deep near-infrared surveys could detect such objects past a redshift of $\sim$20 enabling a probe of the earliest star formation.  A late rebrightening -- most prominent at shorter wavelengths -- is seen about two months after the peak brightness, which is itself  as bright as a superluminous supernova. The ultraviolet spectra during the rebrightening are dominated by the continuum without the broad absorption or emission lines seen in SLSNe or tidal disruption events and the early optical spectra of ASASSN-15lh. Our spectra show no strong hydrogen emission, showing only Ly$\alpha$ absorption near the redshift previously found by optical absorption lines of the presumed host.  The properties of ASASSN-15lh are extreme when compared to either SLSNe or tidal disruption events.

\end{abstract}

%%%%%%%%%% Keywords %%%%%%%%%%%%%%%%%%%%%%%%%%%%%%%%%%%%%%%%%%%%%%%%%%%%%%%%%%%%
%%%%%%%%%%%%%%%%%%%%%%%%%%%%%%%%%%%%%%%%%%%%%%%%%%%%%%%%%%%%%%%%%%%%%%%%%%%%%%%%%
%\clearpage

\section{Introduction\label{intro}}

Superluminous supernovae (SLSNe) are a recently
discovered class of supernovae that are up to 50 times more luminous
than Type Ia supernovae \citep{Quimby_etal_2007, Smith_etal_2006gy, Quimby_etal_2011}.  
Their observed behavior is varied, and the well-studied examples have been further subdivided (see e.g. \citealp{Gal-Yam_2012} for a review).  
Analogous to classical supernova typing, SLSNe I have no
hydrogen in their spectra and they may be related to Type Ic
supernovae \citep{Pastorello_etal_2010}, whereas SLSN II are hydrogen
rich.  Mechanisms that power normal supernovae, such as the
radioactive decay of $^{56}$Ni and gravitational collapse, cannot
reproduce their total integrated energy and light curve rise and fade
rates \citep{Sukhbold_Woosley_2016}.
\citet{Inserra_etal_2013} and \citet{Nicholl_etal_2013, Nicholl_etal_2014} find that
energy deposition by the spin-down of a magnetic neutron star 
(a magnetar; \citealp{Kasen_Bildsten_2010, Woosley_2010,Dessart_etal_2012}) 
can explain the light curve behavior of many SLSNe I, yet
the models are not unique and require parameters tuned to the
observations.  Interaction with pre-expelled circumstellar material is
believed to drive the high energy and emission lines observed for
SLSNe II \citep{Smith_etal_2006gy, Chatzopoulos_etal_2011, Chatzopoulos_etal_2013}.  
For the slowest evolving SLSNe, \citet{Gal-Yam_2012,Kozyreva_etal_2014} suggested a
SLSN R class powered by radioactive decay of several solar masses of
Ni, as was argued for SN~2007bi \citep{Gal-Yam_etal_2009}.  Their
progenitors are predicted to be extremely massive ($\sim$140--260
M$_\odot$) and result in a pair-instability supernova explosion (PISN).  
Objects similar to SN~2007bi but with better data, however, are not consistent with the pair-instability model and suggest magnetars as the likely explosion mechanism \citep{Nicholl_etal_2013}.  
Some SLSNe exhibit double peaks in their light curves which may be a signature of pulsational PISNe \citep{Leloudas_etal_2012,Cooke_etal_2012,Nicholl_Smartt_2016}.
One object, iPTF13ehe, exhibits characteristics of a large radioactive nickel mass and magnetar heating, as well as late time circumstellar interaction \citep{Yan_etal_2015, Wang_etal_2015}.

In addition to their higher optical luminosity, an important difference between SLSNe and SNe Ia is their UV luminosity.  SNe Ia exhibit a sharp drop in flux at wavelengths below 3000 \AA.  This drop has been exploited to identify SNe Ia in the Hubble Deep Field \citep{Riess_etal_2004b} but makes it hard to observe higher redshift SNe in the optical.  It imposes a de-facto redshift limit of $\sim$1.7 beyond which optical filters cover the intrinsically faint UV region.  In contrast, SLSNe show significant UV flux.  SLSNe I do not show the strong metal line blanketing which suppresses the UV flux in other type I SNe.  
  SLSNe II show a very strong rise in flux to shorter wavelengths, similar to the hot photospheres of the classical hydrogen-dominated SNe II.  For the earliest observations of SN~2008es (z=0.205) the wavelength of the peak flux was shortward of the Swift UV observations (rest wavelength of $\sim$1500 \AA; \citealp{Gezari_etal_2009}).  Thus UV observations are important for measuring the total luminosity and constraining the temperature.  The lack of UV data can make it hard to compare the luminosities of SLSNe \citep{Chomiuk_etal_2011}.   From a purely observational standpoint, the high UV flux makes it much easier to detect these SNe at large distances because the UV flux redshifting into the optical bands helps rather than hurts. The current most distant SLSN was discovered at z=3.9 \citep{Cooke_etal_2012}.  Finally, as astrophysical tools, SLSNe can be used as backlights to probe absorption from the interstellar medium similar to quasars and GRBs \citep{Berger_etal_2012}.

While we focus on the SLSN scenario, we will also discuss how the observations compare to those of tidal disruption events (TDEs) --luminous, hot transients resulting from the tidal disruption of a star by a previously inactive galactic nucleus \citep{Rees_1988,Loeb_Ulmer_1997}. 

The luminous ASASSN-15lh \citep{Dong_etal_2016} was discovered by the All Sky Automated Survey for
SuperNovae (ASASSN) and announced as a hydrogen-poor SLSN with rest-frame u-band absolute magnitude of
$-23.5$ (AB system) and extremely blue UV colors \citep{Dong_etal_2015ATEL}.  
The unprecendented luminosity challenged many of the adopted models for SLSNe \citep{Dong_etal_2016}.  
The unknown mechanism powering these explosions and the extreme nature of ASASSN-15lh make multi-wavelength data of great importance to constraining possible models.  In Section \ref{obs} we present the Swift/UVOT and HST observations and briefly outline the calibration and reduction.  In Section \ref{results} we describe the light curve, implied symmetry from the polarimetry, spectral shape, UV spectrum, and the X-ray limits.  In Section \ref{discussion} we discuss the nature of the rebrightening, some similarities and differences to TDEs, and the detectability of such objects to very high redshifts.  We summarize in Section \ref{summary}.  
%%%%%%%%%%%% 
 
\section{Observations of ASASSN-15\lowercase{lh} } \label{obs}

ASASSN-15lh, was discovered by the All Sky Automated Survey for
SuperNovae (ASAS-SN) on UT 2015-06-14.25 (MJD 57187.25) and announced June 16 \citep{Nicholls_etal_2015}.  
It was present near the detection limit of V $\sim$ 17.3
on May 18.32 and not detected on May 15.33 or before to a limit of V$\sim$ 17.3.  UV observations 
with the Swift/Ultraviolet Optical Telescope (UVOT;
\citealp{Roming_etal_2005}) showed it to be extremely blue \citep{Dong_etal_2015ATEL}.  
\citet{Dong_etal_2015ATEL} also reported a spectral classification on
2015 July 8, describing a blue mostly featureless continuum with broad
OII lines similar to hydrogen-poor SLSNe I.  They determined a redshift
of z=0.2326 from MgII absorption lines, implying an absolute magnitude in the rest-frame u band of
$-23.5$ (AB system).  A detailed description of the early UV and optical properties is given by \citet{Dong_etal_2016}.  Fitting a parabola to the V-band data yields the epoch of maximum light in the optical to be 2015 June 05 (MJD 57178.5).

\subsection{Swift Ultraviolet Optical Telescope and X-Ray Telescope Observations}

\begin{figure*} 
\resizebox{16cm}{!}{\includegraphics*{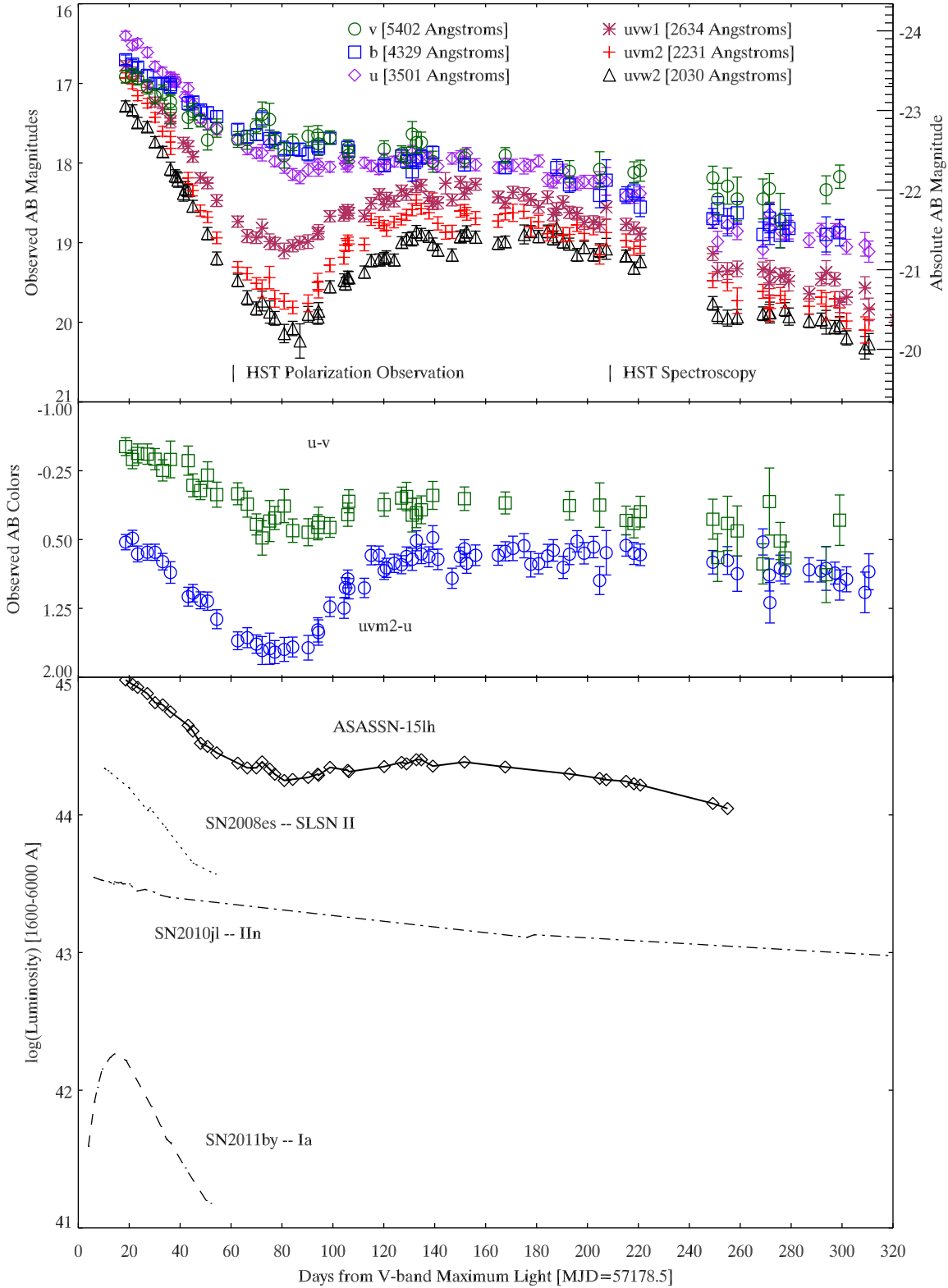}   }
\caption[Results]
        {Top Panel: UVOT light curves of ASASSN-15lh in AB magnitudes.  The right axis gives the absolute magnitude, which is calculated by subtracting a distance modulus of 40.34 from the observed magnitude.  The late time rebrightening is brighter than the m$<-21$ cut-off for superluminous SNe \citep{Gal-Yam_2012}.  Vertical lines indicate the epochs of the HST polarimetric and UV spectroscopic observations.  The x-axis is given in days since the optical maximum \citep{Dong_etal_2016}.
Middle Panel: uvm2-u and u-v colors of ASASSN-15lh.  The SN begins quite blue but reddens with time until the rebrightening during which the SN becomes bluer in the MUV-NUV.  Bottom Panel: The integrated luminosity (in units of ergs s$^{-1}$ ) of ASASSN-15lh between 1600-6000 \AA.  Overplotted is the integrated luminosity of the SLSN II SN~2008es, which is of comparable brightness to the rebrightening event, as well as a bright SN IIn and a normal SN Ia.
 } \label{fig_time}    
\end{figure*}

Observations with Swift began 2015 June 24 00:37:49 (MJD 57197).  The early UVOT data were reported by \citet{Dong_etal_2015}.  Following the announcement of the superluminous nature of ASASSN-15lh \citep{Dong_etal_2015ATEL} we triggered our Swift Guest Investigator programs ``Ultraviolet Properties of Superluminous Supernovae over Ten Billion Years'' (PI: Brown) to obtain UV/optical photometry with Swift/UVOT and ``Late-Time X-Rays from Superluminous Supernovae: How Hard Could it
Be?'' (PI: Quimby).  We have reduced all of the data obtained through  2016 Apr 1.  The reduction utilizes the same routines as the Swift Optical/Ultraviolet Supernova Archive (SOUSA; \citealp{Brown_etal_2014_SOUSA}).  One change is that the time-dependent sensitivity correction of \citet{Breeveld_etal_2011} has been updated\footnote{\url{http://heasarc.gsfc.nasa.gov/docs/heasarc/caldb/swift/docs/uvot/uvotcaldb\_throughput\_03.pdf}}.  No attempt has been made to subtract off the count rates of the underlying galaxy, only expected to be significant at the faintest epochs of our optical filters \citet{Dong_etal_2016}.  We use the revised Swift/Vega system and AB zeropoints of \citet{Breeveld_etal_2011}.
The Swift/UVOT photometry on the Swift/Vega system is given in Table \ref{table_uvot} and displayed in the AB system (for appearance reasons) in Figure \ref{fig_time}.

One might be concerned about the effect of the optical tails of the uvw2 and uvw1 filters (sometimes referred to as the ``red leaks'') which still have $1/1000$ of the peak transmission all the way to 5000 \AA (see \citealp{Breeveld_etal_2011} and \citealp{Brown_etal_2010} for graphical descriptions of the revised filter throughput curves).  To estimate the contribution from the optical tails, we compare the count rates to those expected through synthetic filters similar to uvw2 and uvw1 but cutting off sharply to zero throughput at 2500 and 3300 \AA \citep{Brown_etal_2010}.  Using a blackbody spectrum corresponding to 16,000 K (the hottest estimated below), the counts coming from the redder portion correspond to 12\% and 3 \% of the total in the uvw2 and uvw1 filters, respectively. Using a blackbody spectrum corresponding to 10,770 K (the coolest estimated below), the extra counts coming from the redder portion correspond to 20\% and 5 \% of the total in the uvw2 and uvw1 filters, respectively.  Thus the contribution from the red tails increases as the SN reddens, but never contributes a large fraction of the counts.  The wavelengths beyond which are considered to be part of a red leak (as opposed to just the long wavelength edge of the filter) can be arbitrary and require a spectral model as well as the filter curves to quantify.

Even if the red leak contributed a significant fraction of the observed count rate in the UV filters, the analysis below takes the red leaks into account.  For example, blackbody fits are performed by comparing the observed count rates to thos predicted from blackbody spectra through the whole filter.  Brown et al. (2016, AJ accepted) give a detailed description of issues involved with the conversion of Swift/UVOT broadband measurements into monochromatic flux densities for the generation of spectral energy distributions and bolometric light curves.

\begin{deluxetable*}{ccccccccc}
\tablecaption{Swift/UVOT Photometry of ASASSN-15lh\label{table_uvot}}
\tablehead{\colhead{Name} & \colhead{Filter} & \colhead{MJD} & \colhead{Mag} & \colhead{Mag Error} & \colhead{Upper Limit} & \colhead{Lower Limit} & \colhead{Rate} & \colhead{Rate Error}  \\ 
\colhead{} & \colhead{} & \colhead{(days)} & \colhead{} & \colhead{} & \colhead{(mag)} & \colhead{(mag)} & \colhead{(counts s$^{-1}$)} & \colhead{(counts s$^{-1}$)} } 

%% All data must appear between the \startdata and \enddata commands
\startdata

ASASSN-15lh     & UVW2     & 57197.0974 &  15.555 &   0.066 &  20.993 &  11.085 &   5.371 &   0.327 \\
ASASSN-15lh     & UVW2     & 57199.7898 &  15.606 &   0.067 &  20.940 &  11.085 &   5.125 &   0.314 \\
ASASSN-15lh     & UVM2     & 57197.1005 &  15.240 &   0.062 &  20.358 &  10.555 &   4.405 &   0.252 \\
ASASSN-15lh     & UVM2     & 57199.7948 &  15.316 &   0.062 &  20.408 &  10.555 &   4.106 &   0.235 \\
ASASSN-15lh     & UVW1     & 57197.0939 &  15.265 &   0.060 &  20.519 &  11.145 &   7.410 &   0.408 \\
ASASSN-15lh     & UVW1     & 57199.7838 &  15.344 &   0.061 &  20.414 &  11.145 &   6.890 &   0.384 \\
ASASSN-15lh     & U        & 57197.0950 &  15.382 &   0.056 &  20.287 &  12.051 &  15.252 &   0.784 \\
ASASSN-15lh     & U        & 57199.7854 &  15.500 &   0.061 &  20.274 &  12.051 &  13.682 &   0.764 \\
ASASSN-15lh     & B        & 57197.0957 &  16.833 &   0.060 &  20.704 &  12.832 &   8.145 &   0.454 \\
ASASSN-15lh     & B        & 57199.7866 &  16.888 &   0.061 &  20.712 &  12.830 &   7.738 &   0.438 \\
ASASSN-15lh     & V        & 57197.0993 &  16.925 &   0.080 &  19.537 &  11.609 &   2.433 &   0.178 \\
ASASSN-15lh     & V        & 57199.7923 &  16.903 &   0.081 &  19.480 &  11.609 &   2.481 &   0.185 \\

\enddata

%% Include any \tablenotetext{key}{text}, \tablerefs{ref list},
%% or \tablecomments{text} between the \enddata and 
%% \end{deluxetable} commands

\tablecomments{Magnitudes are given in the UVOT/Vega system for easier comparison with other SNe published in that system (see \citealp{Brown_etal_2014_SOUSA} for a review).  The three $\sigma$ upper limits and lower (saturation) limits come from the exposure and background parameters.  Most of the photometry of ASASSN-15lh is not close to either of these limits, but they are given for completeness and easy comparison with other SN photometry tables from the Swift Optical/Ultraviolet Supernova Archive (SOUSA; \citealp{Brown_etal_2014_SOUSA}).  This table gives a portion of the data, while the full table will be available in machine readable format in the online version.}

%% No \tablerefs 

\end{deluxetable*}

\begin{deluxetable}{cccc}

\tablecaption{0.3 - 10 keV X-ray 3 Sigma Upper limits\label{table_xrt}}
\tablehead{\colhead{MJD} & \colhead{Exposure} & \colhead{Flux Limit} & \colhead{Luminosity Limit} \\ 
\colhead{(days)} & \colhead{(s)} & \colhead{(ergs cm$^{-2}$ s$^{-1}$)} & \colhead{(ergs s$^{-1}$)} } 

%% All data must appear between the \startdata and \enddata commands
\startdata
57197:57476 & 227634  & 1.118e-15 & 1.833e+41 \\
57197:57283 & 53548  & 1.006e-14 & 1.650e+42 \\
57283:57476 & 174086 & 6.257e-15 & 1.026e+42 \\
57197.03 & 2484 & 1.024e-13 & 1.679e+43 \\
57199.75 & 2354 & 1.432e-13 & 2.348e+43 \\
57201.75 & 2267 & 1.122e-13 & 1.840e+43 \\
57205.53 & 2874 & 1.171e-13 & 1.920e+43 \\
57208.60 & 2854 & 8.912e-14 & 1.461e+43 \\
57211.52 & 2667 & 9.538e-14 & 1.564e+43 \\
\enddata

%% Include any \tablenotetext{key}{text}, \tablerefs{ref list},
%% or \tablecomments{text} between the \enddata and 
%% \end{deluxetable} commands

\tablecomments{The MJD column gives the start:stop range of dates for the coadded limits and the midpoint for the single observations.  This table gives a portion of the data, while the full table will be available in machine readable format in the online version.}

%% No \tablerefs indicated

\end{deluxetable}

\begin{deluxetable*}{llccccc}
\tablewidth{0pc}
\tabletypesize{\scriptsize}
\tablecaption{HST Observations on 2015-08-03 \label{table_pol}}
\tablehead{
\colhead{ACS/WFC} & & \colhead{Date of Obs.} \vspace{-0.1cm} 
& \colhead{Exp. Time} & \colhead{MAST Label} & \colhead{Polarization} & \colhead{Position Angle} \\ \vspace{-0.1cm}  
& \colhead{Polaroid} & & & \\ 
\colhead{Filters} & & \colhead{UT} & \colhead{(seconds)} & \colhead{Data Set} & \colhead{(\%)} & \colhead{Degrees} }
\startdata
      & POL0UV   & 17:05:03 & 3$\times$138 & JCVH01040 & \nodata & \nodata \\
F435W & POL120UV & 17:23:06 & 3$\times$138 & JCVH01050 & \nodata & \nodata \\
      & POL60UV  & 17:41:09 & 3$\times$138 & JCVH01060 & 0.60$\pm$0.11 & 23$\pm5^\circ$ \\
\hline
      & POL0V    & 15:32:13 & 3$\times$122 & JCVH01010 & \nodata & \nodata \\
F606W & POL120V  & 15:49:31 & 3$\times$122 & JCVH01020 & \nodata & \nodata \\
      & POL60V   & 16:06:46 & 3$\times$122 & JCVH01030 & 0.78$\pm$0.09 & -11$\pm3^\circ$  \\
\hline
      & POL0V    & 18:35:27 & 3$\times$137 & JCVH01070 & \nodata & \nodata \\
F775W & POL120V  & 18:53:30 & 3$\times$137 & JCVH01080 & \nodata & \nodata \\
      & POL60V   & 19:11:30 & 3$\times$137 & JCVH01090 & 0.63$\pm$0.11 & 30$\pm5^\circ$ \\
\enddata
\end{deluxetable*}

%2015-12-29 

%18:01:07 434	STIS	52X0.1	G430L	2900-5700 2358-4634
%18:14:56 1217	STIS	52X0.1	G230L	1570-3180 1276-2585
%19:27:43 2121	COS	PSA	G140L	1121-2148 911-1746

\begin{deluxetable*}{lrrllll}

\tablecaption{HST Spectroscopy from 2015 Dec 29\label{table_spectra}}

\tablehead{\colhead{Start Time} & \colhead{Exposure} & \colhead{Detector} & \colhead{Slit}  & \colhead{Grism}  & \colhead{Observed Range} & \colhead{Rest-Frame Range} \\ 
\colhead{UT} & \colhead{seconds} & \colhead{} & \colhead{} & \colhead{} & \colhead{Angstroms} & \colhead{Angstroms} } 

%% All data must appear between the \startdata and \enddata commands
\startdata
18:01:07  & 434 & STIS/CCD & 52X0.1 & G430L & 2900-5700 & 2358-4634 \\
18:14:56  & 1217 & STIS/MAMA & 52X0.1 & G230L & 1570-3180 & 1276-2585 \\
19:27:43  & 2121 & COS & PSA & G140L & 1121-2148 & 911-1746 \\

\enddata
\end{deluxetable*}

 \begin{figure} 
\resizebox{8.8cm}{!}{\includegraphics*{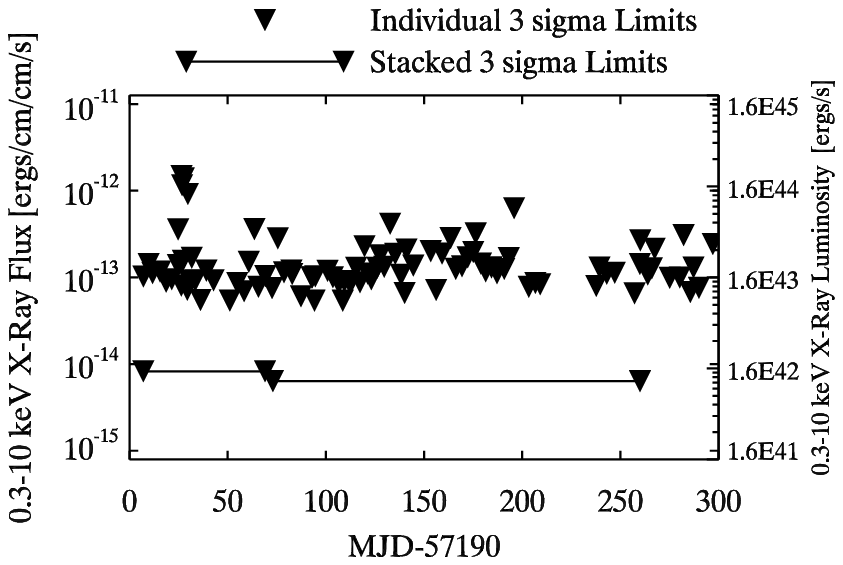}   }
\caption[Results]
        {X-ray upper limits in the 0.3-10 keV range measured from Swift/XRT.  Individual limits are shown as well as the coadded depths during the whole observation sequence, the initial peak, and the rebrightening episode.
 } \label{fig_xrays}    
\end{figure}

Observations with the X-Ray Telescope \citep{Burrows_etal_2005} were obtained simultaneously with UVOT.  
Count rates were extracted as follows: 
Source regions were selected at a 5 pixel radius according to the in-flight calibration of the Swift XRT point spread function \citep{Moretti_etal_2005}.  Similarly, the background region radius was selected with a 10 pixel inner radius and a 100 pixel outer radius. Counts were extracted using the HEASoft tool xselect from each event file to determine total counts in a specified region and exposure time.  Where observations were taken within 14.4 minutes of each other, the data were combined and the limits recalculated. 

Limits for the source were determined using the Bayesian method outlined in \citet{Kraft_etal_1991}, constructing a posterior probability distribution function based on N counts in the source region and B expected background counts. Minimum and maximum source limits were then taken at a given confidence limit. Table \ref{table_xrt} and Figure 
\ref{fig_xrays} provide three sigma limits in unabsorbed flux and luminosity for each of the individual epochs.  It also provides stacked limits for the two periods before and after the rebrightening divided at MJD 57260 and for the sum total of all observations.  
Flux conversions were determined using the HEASARC web tool WebPIMMS. The following constraints were used to compute the conversion factor between count rate and unabsorbed flux: 2.94$\times 10^{10}$ weighted average nH (from HI Column Density Map of \citealp{Dickey_Lockman_1990}), photon index of 2 \citep{Levan_etal_2013}, and the 0.3-10 keV Swift XRT energy range. A conversion factor for count rate to unabsorbed flux was predicted at 3.854$ \times 10^{-11}$ ergs cm$^{-2}$ s$^{-1}$ per count s$^{-1}$.

\subsection{Hubble Space Telescope Imaging Polarimetry}

We requested Director's Discretionary Time observations with the HST to obtain imaging polarimetry (\#14348  PI: Yang).
Our $HST$ ACS/WFC multi-band imaging 
polarimetry of ASASSN-15lh occured approximately t$_p$=59 days 
after the $V$-band maximum light. \\
\\
The observations were taken with three different filters: 
F435W, F606W, and F775W. 
For the F435W filter, it was combined with one of the 
three UV polarized filters: POL0UV, POL60UV, and POL120UV. 
F606W and F775W were combined with one of the 
three visual polarized filters: POL0V, POL60V, and POL120V. 
Table \ref{table_pol} presents a log of the observations. \\
\\
Multiple dithered exposures were taken in each observing 
configuration to allow for drizzling of the images. 
Bright stars and central regions of background galaxies 
in the field of view have been selected as an optimal 
input catalogue for image alignment through 
`tweakreg' in astrodrizzle \citep{Gonzaga_etal_2012}. 

\begin{figure} 
\resizebox{8.8cm}{!}{\includegraphics*{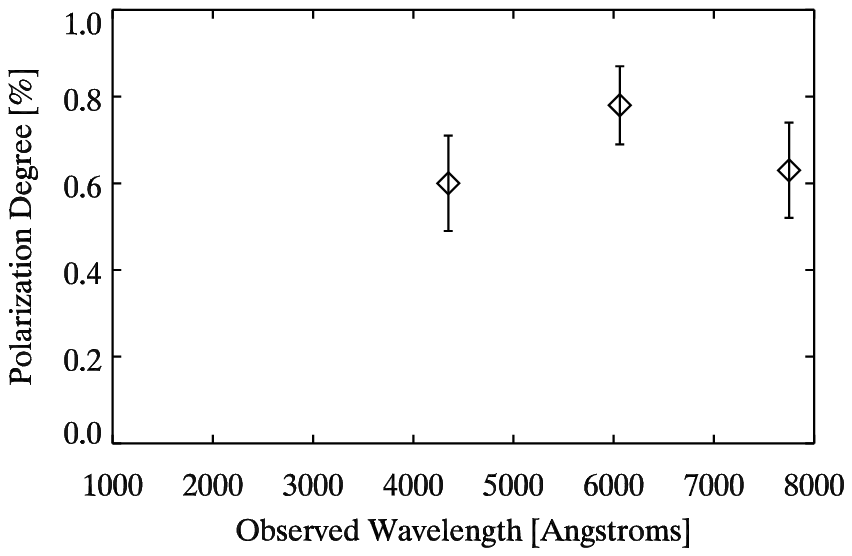}   }
\caption[Results]
        {Wavelength dependence of the polarization measured with HST.
 } \label{fig_pol}    
\end{figure}

Aperture photometry was performed on the drizzled images 
for F435W, F606W, and F775W using an aperture 
radius of 10 pixels (0.5$''$). 
The local background level is estimated with 
a circular annulus of inner radius 20 pixels (1.0$''$) 
and outer radius 30 pixels (1.5$''$) around the source. \\
\\
The calibration of the polarization data is described in more detail in Appendix A.  
The calculated polarization degree and angle for each filter are given in Table \ref{table_pol} and displayed in Figure \ref{fig_pol}.
Not included in the errors are the calibration uncertainties (0.33\% and 4 degrees for F435/POLUV and 0.24\% and 5 degrees for F775W/POLV) from the standard deviation of the measurements of polarized standards as described in Appendix A.

\subsection{Hubble Space Telescope Ultraviolet Spectroscopy}

\begin{figure*} 
\resizebox{16cm}{!}{\includegraphics*{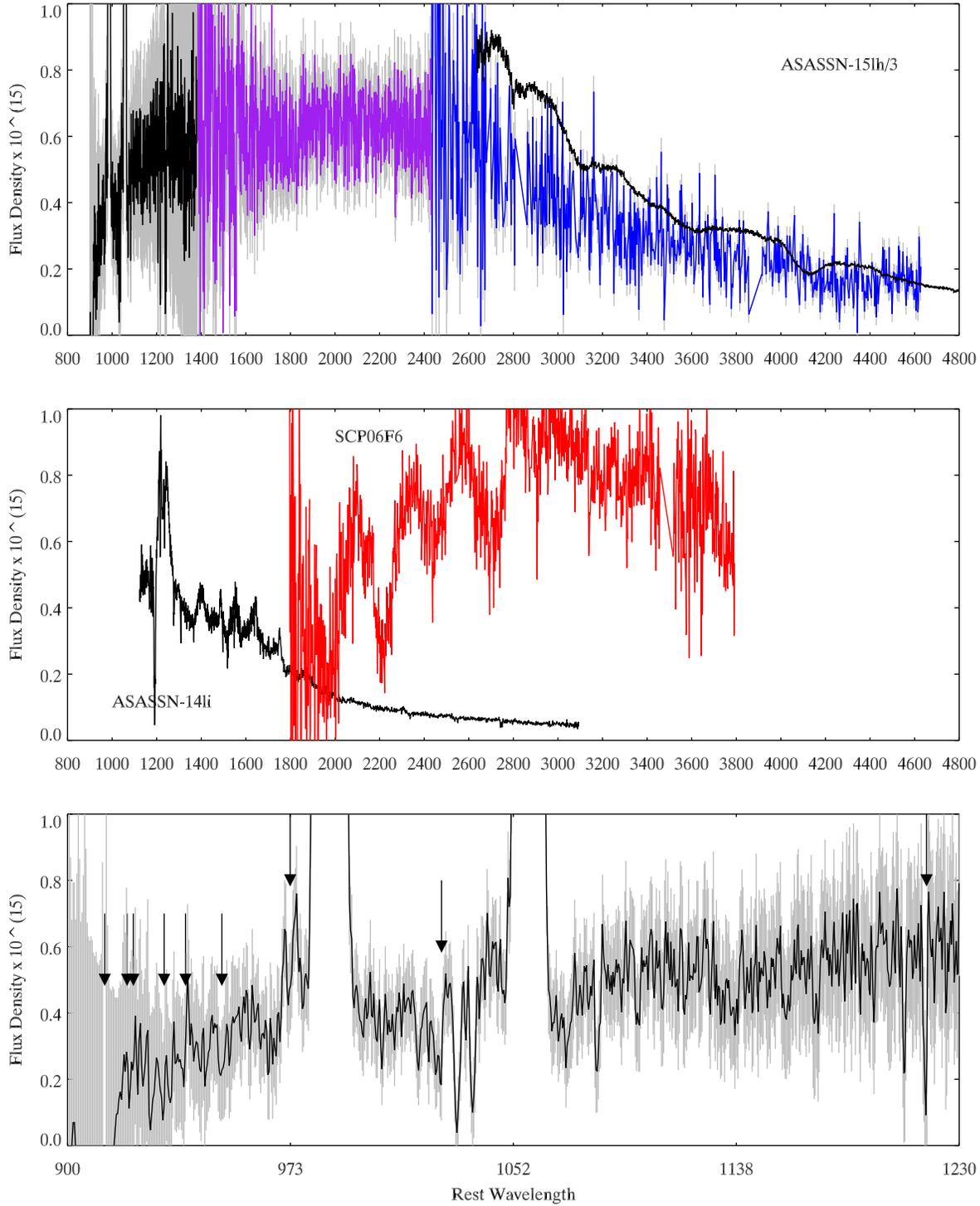}   }
\caption[Results]
        {Top Panel: UV spectra of ASASSN-15lh during its rebrightening phase from COS and STIS (MJD 57386) along with an optical spectrum during its first peak (from SALT on MJD 57203, with flux scaled by a factor of 1/3; \citealp{Dong_etal_2016}).
The grey lines indicate the one sigma flux uncertainty.
Middle Panel: For comparison we show spectra from SLSN SCP06F6 (VLT spectrum on MJD 53873; \citealp{Barbary_etal_2009}) and tidal disruption event ASASSN-14li (from HST on MJD 57034; \citealp{Cenko_etal_2016}).
Bottom Panel: The COS FUV spectrum of ASASSN-15lh during the rebrightening is shown along with the positions of Lyman series transitions.  The grey lines indicate the one sigma flux uncertainty.
 } \label{fig_spectra}    
\end{figure*}

After ASASSN-15lh experienced a significant rebrightening at UV wavelengths, we also requested Director's Discretionary Time observations to obtain a UV spectrum during the rebrightening (\# 14450 PI: Brown).  
STIS observations using the G430L grism occurred on 2015 Dec 29 (see Table \ref{table_spectra} for details).  For these spectra we use the default HST reduction obtained from the Mikulski Archive for Space Telescopes (MAST;\footnote{\url{https://archive.stsci.edu/hst/}}).  We exclude any pixels flagged in the data quality array as well as two single-pixel spikes in the HST/STIS/CCD spectrum.
We also obtained COS observations with the G140L grism with a central wavelength of 1105 \AA. Four different wavelength positions were used.  To improve the S/N from the default reduction, we extracted the spectrum with a smaller (17 pixel tall) extraction box.  
The four wavelength pointings were combined using a weighted average all pixels within a 0.5 \AA~bin.  Because of the wavelength shift, both ends of the spectrum use progressively fewer of the pointings.  The far blue end reaches  $\sim$ 1110  \AA~(though with low sensitivity at the end), while the net flux drops well below the noise before the red end is reached.  Here it is trimmed at 1700 \AA.
The spectra are shown in Figure \ref{fig_spectra}.  Table \ref{table_spectra} details the observation parameters and the rest-frame wavelengths covered by these observations.
%We use a plot of the S/N versus wavelength to determine which wavelength regions to use for each spectrum when combining into a single spectrum.  

%%%%%%%%%%%%%%%%%%%%%%%%%%%%%%%%%%%%%%%%%%%%%%%%%%%%%%%%%%%%%%%%%%%%%%
\clearpage
\section{Results}\label{results}

\subsection{Temporal Behavior\label{lightcurve}}

The ASAS V-band observations show the Swift observations began around the time of maximum light in the optical \citep{Dong_etal_2016}.  The Swift observations fade in all 6 filters, with the decay rate increasing to shorter wavelengths.  This is shown in Figure \ref{fig_time}.  This wavelength-dependent fading and reddening continues to about $50-80$ days after maximum light.  At that point, the optical light curves begin fading more slowly and the UV light curves brighten.  After 60 days of UV brightening, the uvm2-u colors are as blue as they were when Swift first began observing it.  The absolute magnitudes are discussed in \citet{Dong_etal_2016}, with comparisons showing it to be the brightest SN ever discovered.  The rebrightening is by itself as luminous in the UV as SLSNe. The rebrightening is not smooth, but exhibits possible flares or bumps.  We note that this rebrightening appears quite different from the ``double-peaked'' SLSNe which exhibit an early time peak preceding the rise to the primary maximum \citet{Nicholl_Smartt_2016}.  While the rebrightening is obvious in the UV, it only manifests itself as a flattening in the optical decay rate and thus would not have been apparent without observer-frame UV observations.
Thereafter the SN begins a slow decay in the optical, steeper in the UV filters.  There may be an additional bump or leveling off in UVOT's shortest wavelength uvw2 filter at 260 days after maximum light.

To compute an integrated luminosity, we create an SED in as discussed in Brown et al. (2016, submitted).  
In summary, we begin the conversion from count rates to flux density at the Vega effective wavelengths using the average values from \citet{Poole_etal_2008}.  We linearly interpolate between the points and vary the flux density at the effective wavelength points until the spectrophotometric count rates from the SED match the observed count rates.  
Using the SEDs, we integrate the luminosity within the UVOT range (1600-6000 \AA) and plot the luminosity evolution in the bottom panel of Figure \ref{fig_time}.  Also plotted is the integrated luminosity of the UV-bright SLSN II SN~2008es \citep{Gezari_etal_2009}.  The late rebrightening of ASASSN-15lh is as bright as SN~2008es.

\subsection{Geometry of the Explosion}\label{geometry}

The polarization degree is plotted as a function of wavelength in  Figure \ref{fig_pol}. 
The flat polarization spectrum is inconsistent with that expected from interstellar dust in the Milky Way or the host galaxy.  However, the shape of the polarization spectrum can vary depending on the dust properties. The Milky Way extinction along this line of sight is low, A$_V\sim0.1$, and the internal extinction from the host galaxy is inferred to be low based on the very blue SED of ASASSN-15lh.  An empirical upper bound of 9.0\% mag$^{-1} \times$ E(B-V) on interstellar polarization due to Milky Way dust  was derived by \citet{Serkowski_etal_1975}. For A$_V$ = 0.1, we expect the degree of polarization due to Milky Way dust to be less than 0.29 \%, which is smaller than the observed polarization.  Most of the observed polarization is likely from the SN itself, but at the least the measured polarization represents an upper limit on the intrinsic polarization.  The 0.6-0.8\% polarization is high for a SN compared to e.g. Type Ia SNe (continuum or line polarization $\leq$ 0.3\% \citealp{Wang_Wheeler_2008}), but not that large, implying the emitting photosphere has a small asymmetry on the sky. It is comparable to that recently measured for the SLSN I LSQ14mo \citep{Leloudas_etal_2015}.  It is also comparable to the core of SN~IIP~2004dj \citep{Leonard_etal_2006} Having only one epoch prevents us from measuring the time dependence and the change in shape as has been observed in suspected jet-driven SNe 2006aj and 2008D \citep{Gorosabel_etal_2006, Maund_etal_2007_06aj, Maund_etal_2009, Gorosabel_etal_2010}. 

Broad-band polarimetry does not yield much detailed information on line polarization.  The consistency of the polarization measurements suggests that the OII line (seen in absorption at $\sim4100$ \AA~by \citealp{Dong_etal_2016} and redshifted into the F606 filter bandpass), does not have a significantly higher polarization than the continuum which dominates the redder filters.   

The geometry (or even the exact nature) of the energy injection from a magnetar is not often addressed.  \citet{Bucciantini_etal_2009} used axisymmetric magnetohydrodynamic simulations to study the interaction.  The formation of a polar jet is very asymmetric, yet for a GRB this jet punches through the SN explosion without transferring much energy.  If a jet stalls early, one might also see roughly spherical ejecta from an intrinsically asymmetric explosion, provided the surrounding material is roughly spherical \citep{Maund_etal_2009}.   \citep{Soker_2016} suggests it is the jets themselves which power SLSNe.  Such models should be explored further to determine the effect the jet might have on the observed luminosity and asphericity of the SN ejecta.  

\subsection{Early Spectral Shape\label{sed}}

\begin{figure} 
\resizebox{8.8cm}{!}{\includegraphics*{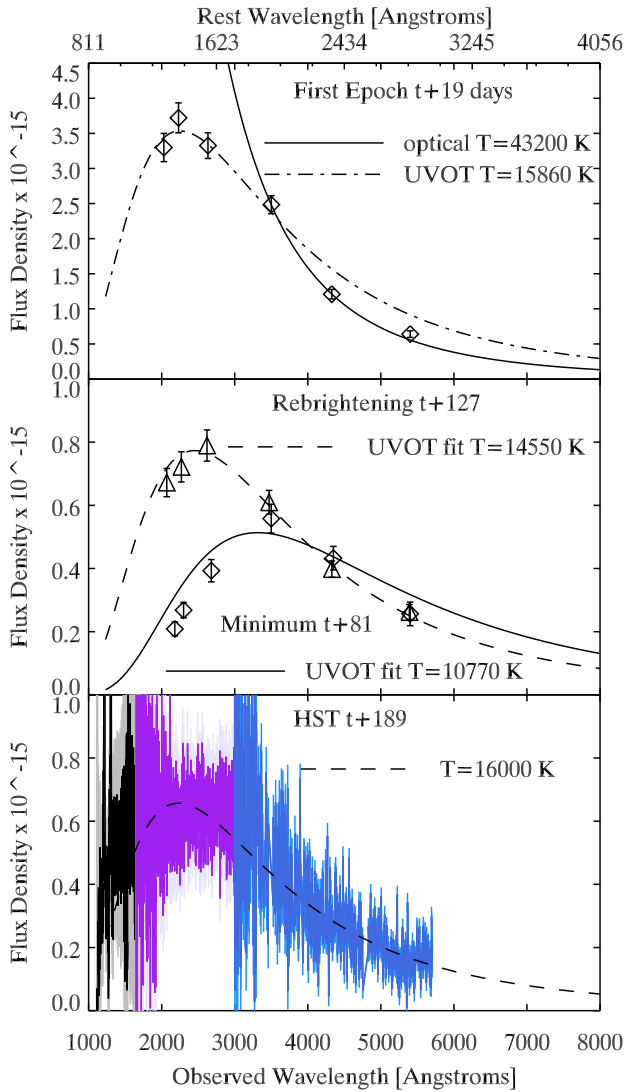}   }
\caption[Results]
        {Top Panel: UVOT SEDs (in units of ergs s$^{-1}$ cm$^{-2}$ \AA$^{-1}$) from the first epoch (MJD=57197) compared to blackbody fits derived from the optical photometry (u,b,v; plotted with a solid line) and all six UVOT filters (plotted with a dotted-dashed line).  
The flux conversion is spectrum dependent--we plot the UVOT flux densities using conversion factors for the six-filter fit. 
 %Since the flux conversion is spectrum dependent, we plot two versions of the UVOT flux densities -- the triangle points are from the optical fit, and the diamond points are from the six-filter fit.  Only the uvw1 and u flux densities differ by more than 1\%.  
Second Panel: SEDS from the light curve minimum  (MJD=57259) and the first peak of the rebrightening  (MJD=57305) compared to 6-filter blackbody fits.   The flux conversions utilize the best-fit blackbody spectra.
Bottom Panel: HST spectra from COS and STIS (MJD 57386) compared to a 16,000 K blackbody.  
 } \label{fig_sed}    
\end{figure}

As apparent from the photometric colors and the SEDs created earlier, the spectrum rises strongly in the near-UV and then peaks in the uvm2 filter (rest-frame wavelength 1800 \AA).  
In the top panel of Figure \ref{fig_sed} we show blackbody fits (in the rest-frame) using all six UVOT filters or just the three optical ($ubv$) filters for the first epoch.  Since the conversion from observed count rates to flux density is spectrum dependent \citep{Poole_etal_2008,Brown_etal_2010}, we use the six-filter blackbody fit spectra to make the conversion for plotting purposes only.  Using the hotter three-filter blackbody spectrum would reduce the uvw1 flux density by 10\% and the u flux by 5\% because so many counts would be detected from blue edges of the filters.  Neither blackbody fits the data well.  The six-filter fit overestimates the optical and underestimates the UV flux.  The extrapolation of the optical fit drastically overestimates the UV flux.  Though line blanketing in the UV is much smaller than observed in radioactively-powered SNe \citep{Quimby_etal_2011}, line-blanketing depression of the UV flux is seen in hydrogen-dominated SNe II as the SN cools (e.g. \citealp{Brown_etal_2007_05cs,Dessart_etal_2008}).  Thus determinations of the temperature using UV photometry can be incorrect \citep{Valenti_etal_2016} and we give only estimates based on the observed spectral shape.  We cannot tell from photometry alone whether the the early spectrum is a blackbody with UV absorption or what its origin is.

As the SN fades, the UV dominates the luminosity at all times, even at its reddest point during the flux minimum.  Note that the magnitudes in Figure 1 are given relative to the AB standard, which has a constant flux density in frequency, and thus a very blue spectral shape.  Thus the colors still appear ``red'' or positive when comparing the UV to the optical.  Best-fit blackbodies SEDs from the minimum and at the peak of the rebrightening are shown in the middle panel of Figure \ref{fig_sed}.  The bottom panel of Figure \ref{fig_sed} shows the HST spectra, confirming a similar spectral shape and showing that the UV luminosity comes from a hot continuum rather than line emission.

\subsection{UV Spectrum During the Rebrightening\label{spectrum}}

Our HST spectra of ASASSN-15lh during the rebrightening shows that the UV luminosity is dominated by a continuum rather than line emission.
Compared to most SN spectra (see e.g. \citealp{Filippenko_1997}), the spectra  appear featureless, at least lacking the broad absorption and emission lines usually seen.  This is shown in Figure \ref{fig_spectra}.  The flux is dominated by a continuum broadly peaking at an observed wavelength of 2500 \AA.  The COS spectrum does show narrow hydrogen absorption from the Lyman series and a strong cutoff near that expected for the Lyman break.  The spectra also feature other narrow absorption lines which will be discussed in Cooke et al. (2016, in preparation).  Luminous, featureless spectra are ideal for the study of intervening interstellar, circumgalactic, and intergalactic material.  Events like this will be very powerful tools to study the z$>$6 universe as they are brighter than galaxies at that epoch.

\subsection{X-Ray Limits\label{sec_xray}}

Because of the higher redshift, the X-ray limits (given in Table \ref{table_xrt} and plotted in Figure \ref{fig_xrays}) are not deep enough to have detected SNe comparable to relatively nearby circumstellar-interaction powered SNe IIn which have been detected, such as  SN~2005ip ( $\sim 1.5 \times 10^{41}$ ergs s$^{-1}$; \citealp{Katsuda_etal_2014}), SN~2010jl ( $\sim$$10^{42}$ ergs s$^{-1}$; \citealp{Ofek_etal_2014,Chandra_etal_2015}),  or SN~2006jd ( $\sim 3 \times 10^{41}$ ergs s$^{-1}$; \citealp{Chandra_etal_2012_06jd}).  
Type IIP SNe are fainter than $\sim 10^{42}$ ergs s$^{-1}$ \citep{Dwarkadas_2014}.
These limits are comparable to the limits placed on other SLSNe (e.g. \citealp{Quimby_etal_2011}).  These limits are two orders of magnitudes more sensitive than the possible detection of SCP06F6 \citep{Levan_etal_2013}.  Because of the high sampling it is unlikely that we missed a such a bright X-ray transient in ASASSN-15lh.

\citet{Metzger_etal_2015}  predicted a late onset of X-ray emission ($L_X\sim10^{42}-10^{44}$) from an ``ionization breakout.''  We can exclude the bright end of this range with our well-sampled observations, and our summed limits rule out any long duration emission down to the low end of that range (see also \citealp{Margutti_2015lh,Godoy_etal_2016}).  Since ASASSN-15lh did not behave normally at late times in the optical/UV, that would also need to be taken into account in ruling out or creating new models.

\section{Discussion}\label{discussion}

\subsection{The Nature of the Rebrightening}\label{discussrebrightening}

The unprecedented peak brightness was noted by \citet{Dong_etal_2016} to pose problems for most supernova explosion models.  The radioactive decay of $^{56}$Ni is clearly ruled out \citep{Kozyreva_etal_2016, Sukhbold_Woosley_2016}. Theorists have risen to the challenge to explain how the peak brightness could be obtained, with a magnetar being the favored solution \citep{Metzger_etal_2015,Bersten_etal_2016, Sukhbold_Woosley_2016,Dai_etal_2016} though \citet{Chatzopoulos_etal_2016} match the peak with models invoking only CSM interaction or with CSM interaction dominating the early phase.  

In addition to its peak luminosity, ASASSN-15lh was also unusual in displaying a secondary peak in the UV light curve.
Some models for the explosion of massive stars have qualitatively similar late-time peaks in the bolometric luminosity due to the radioactive decay of $^{56}$Ni \citep{Smidt_etal_2014,Smidt_etal_2015}, though Ni and other iron-peak elements in the ejecta would lead to strong line blanketing and a UV flux deficit.  
SLSN I iPTF13ehe also exhibited a late-time excess in the r-band which was accompanied by H-alpha emission from likely circumstellar interaction \citep{Yan_etal_2015, Wang_etal_2015}.  Whether the excess was a flattening or a rebrightening is unclear due to the photometric sampling.  
Optical spectroscopy during the rebrightening of ASASSN-15lh does not show evidence of broad H-alpha  \citep{Milisavljevic_etal_2015} nor do we see strong or broad Lyman alpha emission in our UV spectra which would be expected from interaction with H-rich material.  
Some hydrogen may be present in the ejecta and/or interaction, but ionized because of the high temperatures (the narrow hydrogen absorption we see in Fig \ref{fig_spectra} could be farther out in an ejected shell or the host galaxy).  The interaction could be with H-poor material, a model proposed for the luminous peaks of H-poor SLSNe I \citep{Chatzopoulos_Wheeler_2012_shells, Sorokina_etal_2015}.  
Based on the velocity of a broad line in the early spectra (FWHM $\sim$10,000 km s$^{-1}$; \citealp{Dong_etal_2016}) the start of the rebrightening would suggest material located about 8 $\times 10^{15}$ cm away, and the duration of the rise would suggest a radial extent of 4 $\times 10^{15}$ cm.    The details would be model dependent.

Recently, \citet{Chatzopoulos_etal_2016} proposed several hybrid models which could explain what they describe as a UV-bright plateau in the bolometric luminosity.  They explain that primary peak and plateau as a combination of magnetar and circumstellar interaction.  Reasonable fits to the whole bolometric light curve reproduce the early peak and late plateau with forward shock emission/magnetar and magnetar, respectively (labeled CSM0\_A by \citealp{Chatzopoulos_etal_2016}), magnetar/forward shock and reverse shock emission (CSM0\_B), forward shock and magnetar (CSM2\_A), and forward shock and reverse shock from a pulsational pair-instability SN (CSM0).  
Some of these different scenarios from \citet{Chatzopoulos_etal_2016} can explain a flattening off, for example with a fading circumstellar interaction whose flux drops below that of a broad magnetar-powered bump.  However, in all cases, the component resulting in the late-time plateau was significant at early times as well.
Considering the UV light curves separate from the bolometric luminosity, however, provides an additional constraint pointing to distinct phases likely dominated by different mechanisms.  The rapid UV fading followed by a rebrightening suggests the second component is turning on at late times, rather than just being revealed after the fading of the first component.  
 \citet{Gilkis_etal_2015} suggest that late or prolonged accretion onto the remnant black hole could power such SLSNe in the jet feedback mechanism scenario.  

The late-time UV spectra lack the broad features often seen in the rest-frame UV spectra of SLSNe \citep{Barbary_etal_2009,Quimby_etal_2011} or even in the earlier optical spectra of ASASSN-15lh \citep{Dong_etal_2016}.
If magnetars are the correct interpretation for most SLSNe, then it is unlikely to explain the rebrightening phase for ASASSN-15lh.  
The difference in the early and late phases of ASASSN-15lh argues for different emission mechanisms.

\subsection{A Tidal Disruption Event?}\label{tde}

While we have focused our discussion on SLSN models (the preferred interpretation of \citealp{Dong_etal_2016} due to spectroscopic comparisons with other SLSNe), the projected position of ASASSN-15lh coincides with the nucleus of the presumed host galaxy and shares some similarities with TDEs and theoretical predictions for such.  \citet{Strubbe_Quataert_2009} show that a late-time flattening or even rebrightening can result from sub-Eddington fall-back accretion, though this has not been previously observed.   
Additionally, there are smaller scale chromatic and achromatic wiggles superimposed on the broad rebrightening which may be from density inhomogeneities in the circumstellar material (if the rebrightening is interaction driven), activity in a central engine, or irregularities in the accretion (if a TDE).

A TDE is a very asymmetric effect, with the emission thought to arise from the accretion disk itself or reprocessed by external material (see \citealp{Holoien_etal_2016} for a discussion of the relevant distances for three TDEs observed with Swift).  However, the expected or observed polarization signal from a TDE has not been well studied.  \citet{Wiersema_etal_2012} report a 7.4 $\pm 3.5 \%$ linear polarization in Swift J164449.3+573451, but it is thought to be a beamed jet which outshines the light from the stellar disruption \citep{Bloom_etal_2011,Burrows_etal_2011}.    
If ASASSN-15lh is a TDE, the emitting region must have a small asymmetry on the sky to be consistent with our low measured polarization.  While the required alignment might be rare, the exceptional characteristics of ASASSN-15lh do not encourage ruling out extreme (but allowed) parameters. The measured polarization does seem more like a supernova.

Compared to well-observed TDEs, ASASSN-15lh lacks the broad emission features often seen including hydrogen \citep{Cenko_etal_2016,Arcavi_etal_2014}, helium \citep{Gezari_etal_2012,Holoien_etal_2016}, and higher ionization emission lines \citep{Cenko_etal_2016}.   
\citet{Strubbe_Quataert_2011} predict a featureless optical/near-UV spectrum with absorption lines below 2000 \AA~due to the high temperatures and low densities.

\citet{Dong_etal_2016} and \citet{Godoy_etal_2016} fit blackbody curves for the whole light curve of ASASSN-15lh and use the early cooling and expansion of the blackbody fits as an argument for a supernova event.  The early temperatures, however, were extrapolated, as there were no multiwavelength data during the rise to maximum luminosity.  Their blackbody fits show a cooling through the decline and then a strong reheating until ASASSN-15lh begins to fade again.  The inferred radius rises slightly during the peak and decline and then drops during the rebrightening phase \citep{Godoy_etal_2016}.  This is interpreted to result from the photosphere receding as the outer layers expand and become optically thin. 
 \citet{Godoy_etal_2016} contrast the luminosity, radius, and temperature evolution with three well-observed TDEs to bolster the SLSN interpretation.  However, ASASSN-15lh looks as different from SLSNe (e.g. Figures 5 and 6 from \citealp{Inserra_etal_2013} and Figure 4 from \citealp{Dong_etal_2016}) as it does from TDEs. In particular, the TDE ASASSN-14ae \citep{Holoien_etal_2014} shows a cooling and reheating evolution qualitatively similar to ASASSN-15lh and much different from the monotonic cooling seen in most SLSNe \citep{Inserra_etal_2013}.  The large variety seen within SLSNe and TDEs make it hard to argue that ASASSN-15lh is an extreme example of one and rule out the other when it does not follow either very well.

\subsection{Detectability at High Redshift}\label{highz}

Because ASASSN-15lh had such a bright peak luminosity and remained bright for so long, it is of interest to know how far away such an object could be detected by current or future observatories.  As noted in the introduction, SN explosions are tracers of star formation and SNe or TDEs could be used as backlights to study intervening systems in absorption.  Because of their high luminosity and relatively featureless continua, SLSNe can be used as backlights to study the local environment, the host galaxy ISM, and circumgalactic medium and the IGM.  These can then be studied in emission after the object fades.
Estimating the brightness of objects at higher redshifts is often difficult because of the difference between the rest-frame and observer-frame wavelengths and the limited wavelength range of  observations.  The UV/optical observations of ASASSN-15lh, by covering shorter rest-frame (and observer-frame) wavelengths than usual ground-based observations, give us the amount of flux at those important wavelengths.

\begin{figure*} 
\plottwo{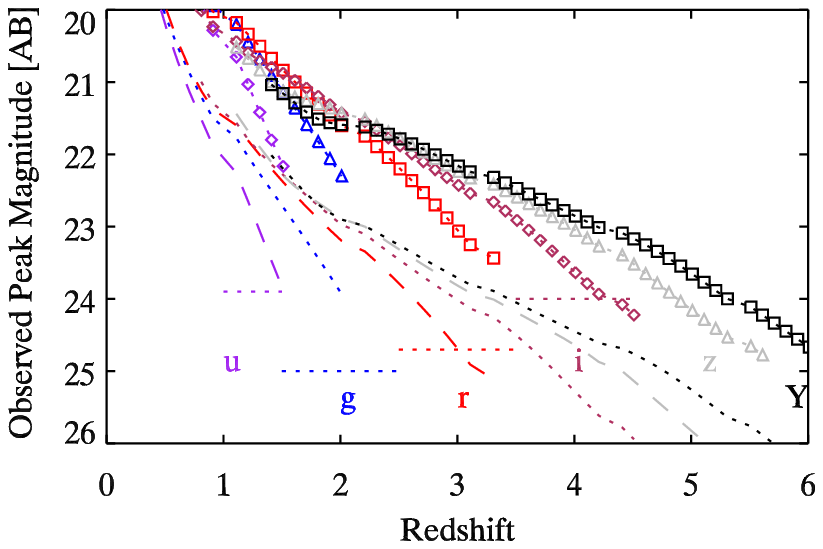} {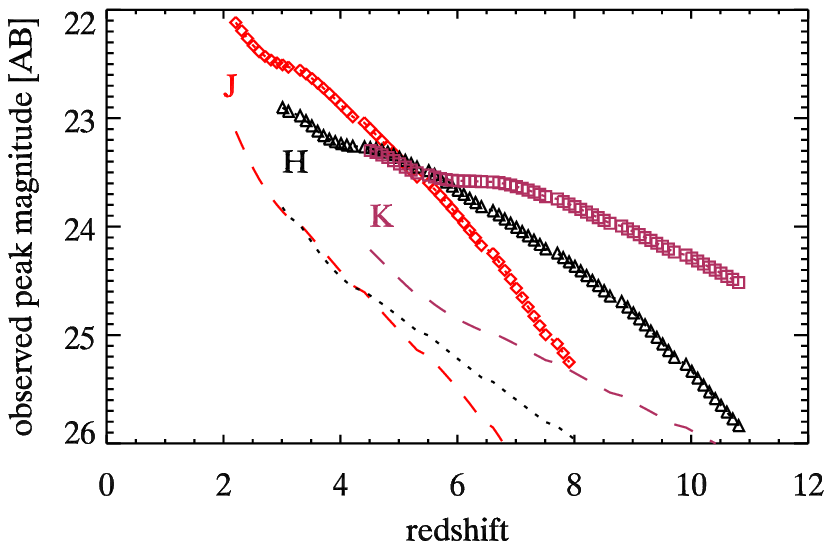} 
\caption[Results]
        {Left Panel: Predicted observed magnitudes (in the LSST filter system) for ASASSN-15lh near peak (symbols) and during the rebrightening (dashed and dotted lines) if it were observed at different redshifts.  The high UV luminosity is redshifted into the optical bands and detectable to redshifts of z$\sim$4 with LSST in single visits.  This includes just the contribution observed by UVOT (rest frame 1300-4900 \AA).  Right Panel: Predicted observed peak magnitudes of the ASASSN-15lh SED near peak (symbols) and during the rebrightening (dashed and dotted lines) redshifted into the NIR. 
 } \label{fig_redshift}    
\end{figure*}

To predict the observability we use the SED (not the best fit blackbody spectra) from ASASSN-15lh from the first epoch as shown in Figure \ref{fig_sed} as well as near the peak of the rebrightening.
The observer-frame SED wavelengths are then shifted into the rest frame using a redshift of z=0.2326 \citep{Dong_etal_2016}.  The flux is increased by a factor of 1+z to account for flux dilution and scaled to a luminosity distance of 1171 Mpc \citep{Dong_etal_2016}.  This gives us a distance-corrected spectral energy distribution in the rest-frame.  For a grid of redshift values, we then redshift this spectrum and correct for flux dilution and the luminosity distance (using lumdist.pro\footnote{http://idlastro.gsfc.nasa.gov/ftp/pro/astro/lumdist.pro} with a standard cosmology with H$_0$=70, $\Lambda=0.7, \Omega_m=0.3$).  We calculate spectrophotometry using these observer-frame SEDs based on filter curves from the Large Synoptic Survey Telescope (LSST) in the $u,g,r,i,z,Y$ and UKIRT for $J,H,K$ \citep{Hewett_etal_2006} through the SNANA program \citep{Kessler_etal_2009_SNANA} with zeropoints computed on the AB system.  The resulting curves of observed AB magnitudes as a function of redshift are displayed in Figure \ref{fig_redshift}.

Although ASASSN-15lh is placed at increasing larger distances/redshifts, the shifting of the bright UV-flux into the observer-frame optical bandpasses keeps the magnitudes from dropping off as quickly.  It would likely be detectable in the bluer optical bands to larger distances than shown, but we cut off at the observed short-wavelength edge of the UVOT observations and do not know how bright it is in the far-UV.  Nevertheless, the redshifted mid-UV flux is detectable in r and i out to a redshift of z$\sim$4 at the single visit depths of LSST.  K-band observations down to K$_{AB}\sim25$ would be able to detect it to z$\sim$12.  
\citet{Tanaka_etal_2012} and \citet{Tanaka_etal_2013} did a similar study using theoretical models and SED matches to the SLSN II 2008es and also estimated rates as a function of redshift.  Identifying these transients at high redshifts would be complicated by the broad, time-dilated light curves, but there are multiple progenitor channels which are certainly bright enough.  Regardless of the nature of ASASSN-15lh, high redshift counterparts would be useful backlights of intervening material.  Using them to understand star formation or black hole evolution would require better ways of distinguishing SLSNe from TDEs.

%%%%%%%%%%%%

\section{Summary}\label{summary}
%%%%%%%%%%%%%%%%%%%%%%%%%%%%%%%%%%%%%%%%%%%%%%%%%%%% 

We have presented UV/optical photometry of ASASSN-15lh spanning nearly three hundred days from Swift/UVOT with only X-ray limits from XRT over the same period.  We observe a remarkable UV rebrightening not seen in previous SLSNe or TDEs. We show from  a single epoch of HST multi-band polarimetry that the emission region could have been only mildly asymmetric as projected on the sky.  HST UV spectra during the rebrightening exhibit a lack of broad hydrogen emission (or any broad emission or absorption features) also dissimilar to SLSNe or TDEs.  We demonstrated that the persistently strong UV flux could be detectable at high redshifts using already planned exposures with the LSST.
By providing this unique UV, X-ray, and polarimetric data we hope to allow others to better constrain models for the progenitor system and explosion to determine more conclusively the nature of this enigmatic object.  A successful model needs to result in a high UV/optical luminosity, low optical polarization, a UV-dominated rebrightening without H or He emission, and no bright X-rays.

%%%%%%%%%%%%%%%%%%%%%%%%%%%%%%%%%%%%%%%%%%%%%%%%%%%%%%%%%%%%%%%%%%%%%%%%%%%%%%%%%

\acknowledgements

We thank the HST director for approving the DDT requests.  
We thank Matt McMaster and Dean Hines for helping with the calibration of the ACS/WFC polarizers.  Based on observations made with the NASA/ESA Hubble Space Telescope, obtained from the data archive at the Space Telescope Science Institute. STScI is operated by the Association of Universities for Research in Astronomy, Inc. under NASA contract NASA 5-26555.  These observations are associated with programs \#14348 and \#14450.  Support for this work was provided by NASA through grant number HST-GO-14450.001-A from the Space Telescope Science Institute, which is operated by AURA, Inc., under NASA contract NAS 5-26555.  JCW was supported by STSCi by STScI grant HST-AR-13276.02-A.
This work is supported by the Swift GI program through grant NNX15AR41G.
The Swift Optical/Ultraviolet Supernova Archive (SOUSA) is supported by NASA's Astrophysics Data Analysis Program through grant NNX13AF35G.
This work made use of public data in the {\it Swift} data
archive from observations requested by several others (PIs: Dong, Godoy, Holoien, Leloudas, Jonker).
This research has made use of NASA's Astrophysics Data System Bibliographic Services.

\bibliographystyle{apj}

\bibliography{bibtex}

\appendix

\subsection{Computing the degree and orientation of linear polarization}

The Stokes vectors and associated errors were 
calculated by following the case of three polarizers 
described by \citet{Sparks_Axon_1999}. 
Neither the ACS visible or UV polarizers are ideal. 
Significant instrumental polarization likely 
comes from the M3 and IM3 mirrors \citep{Biretta_etal_2004}. 
Correction factors C$(CCD, POLnXX, spectral \ filter,n)$ 
in equation \ref{eqn_1} are applied and investigated in 
different calibration proposals: 9586 (P.I. W.B. Sparks), 
9661 $\&$ 10055 (P.I. J. Biretta), and 13964 (P.I. M. McMaster). 
These included observations of unpolarized and polarized stars. \\
\\
HRC has been unavailable since January 2007, 
and all the observations and involved calibrations in the 
most recent polarimetric calibration proposal 13964 
are only for the usage of WFC. 
The polarimetry measurements for the same targets reproduced from 
the on-orbit calibration runs vary on the order of $1\sim2\%$, 
which is most likely due to instrumental effects at the $1-2\%$ level
and a few degrees in the position angle \citep{Sparks_etal_2008}. 
Importantly, the optical chain containing WFC with the filters 
F435W/POLUV has never been calibrated before. 
By applying correction factors to each combination of wavelength 
filter and polarization filter sets \citep{Cracraft_Sparks_2007}, 
one can correct the throughput for each polarizer, thus
removing the instrumental polarization to 
obtain the true polarized level of the source. 
Correction factors for F606W/POLV have been well characterized by 
\citet{Cracraft_Sparks_2007} and reach a precision down to 
$\sigma_P \% \approx 0.3\%$ and $\sigma_{\theta} \sim 3^{\circ}$. 
The correction factors for F775W/POLV have not been tested by 
any other observations until program 13964. \\
\\
We calibrated the throughput of F435W/POLUV and F775W/POLV 
filter sets based on the most recent calibration program 13964. 
The correction factors reproduce the polarization degrees 
of the polarized standard: Vela1-81 measured from the observations 
13964 are: $5.89 \pm 0.09\%$, $6.02 \pm0.09\%$,$5.40 \pm 0.09\%$ 
among three rolling angles, for a mean of $5.77\%$ for F435W, 
compared to the published value 6.1 \% \citep{Whittet_etal_1992}. 
The measured position angles are $6^\circ$, $1^\circ$, $-2^\circ$, 
compared to a published value $1^\circ$  \citep{Whittet_etal_1992}. 
Measurements of the unpolarized standard EGGR-247 give the 
polarization degree to be $0.70 \pm 0.02\%$ for F435/POLUV.

For F775W/POLV, we get $5.42 \pm 0.07\%$, $5.10 \pm 0.07\%$, 
$5.14 \pm 0.07\%$, 
for a mean of $5.21\%$ compared to the published value $6.29\%$  \citep{Whittet_etal_1992}. 
The position angles are $4^\circ$, $3^\circ$, and $8^\circ$, 
compared to $-1^\circ$  \citep{Whittet_etal_1992}. 
The measurement on EGGR-247 gives the polarization degree $0.38 \pm 0.07\%$. 
The correction factors for F775W/POLV listed in Table 1 from 
\citet{Cracraft_Sparks_2007} reproduce: 
$6.12 \pm 0.07\%$, $4.47 \pm 0.07\%$, $5.04 \pm 0.07\%$ 
among the three roll angles, 
for a mean of $5.21\%$ compared to the published value $6.29\%$  \citep{Whittet_etal_1992}. \\
\\
We found that the previously published correction factors for 
F775W/POLV  \citep{Whittet_etal_1992} fail to produce consistent degree of polarizations 
among the observations of the same polarized standard 
with three different roll angles. 
Moreover, the mean level of the polarization degree for Vela1-81 
is measured to have decreased by $\sim1.1\%$. 
Vela1-81 is an OB star \citep{Muzzio_Orsatti_1977}, and the most recent published polarimetry measurement 
is from 1987 \citep{Whittet_etal_1992}. Because many OB stars, especially OB supergiants and Be stars are polarimetric variables \citep{Bjorkman_1994}, we cannot exclude the possibility that the degree of polarization of OB supergiant Vela181 has changed in the last three decades.\\
\\
Our calibration reproduced the polarization to be consistent 
among the three roll angles, with uncertainties: 
$\sigma_F \% \approx 0.33\%$ and $\sigma_{\theta} \sim 4^{\circ}$ 
for F435W/POLUV, 
$\sigma_F \% \approx 0.24\%$ and $\sigma_{\theta} \sim 5^{\circ}$
for F775W/POLV. 
Since F606W/POLV has been done at a single roll angle in 13964, 
and which has been tested by other observing runs, 
we use the correction factors for F606W/POLV listed in 
\citet{Cracraft_Sparks_2007}. 
Table \ref{table_correction} shows the correction factors we used as follows:\\

\begin{equation}
r(n) = C(CCD, POLnXX, spectral \ filter,n) \  r_{obs}(n)'
\label{eqn_1}
\end{equation}
Stokes vectors in the different bands are computed 
for the target using the following equations. 

%\begin{equation}\label{eqn_2}
%\begin{align*}\label{eqn_2}
%$I=\bigg{(} \frac{2}{3} \bigg{)} [r(0) + r(60) + r(120)] $\\
%$Q=\bigg{(} \frac{2}{3} \bigg{)} [2r(0) - r(60) - r(120)]$ \\
%$U=\bigg{(} \frac{2}{\sqrt{3}} \bigg{)} [r(60) - r(120)]$
$I=( \frac{2}{3} ) [r(0) + r(60) + r(120)] $\\
$Q=( \frac{2}{3} ) [2r(0) - r(60) - r(120)]$ \\
$U=( \frac{2}{\sqrt{3}} ) [r(60) - r(120)]$
%\end{align*}
%\end{equation}

The cross-polarization leakage is insignificant 
for POLVIS filters \citep{Biretta_etal_2004}. 
Then F$\%$ is calculated using the Stokes vectors. 
These corrections together with the calibration of the 
source count rates vectorially remove the instrumental 
polarization of the WFC. 
The position angle (P.A.) is calculated using the Stokes 
vectors and the roll angle of the $HST$ spacecraft 
(PA$\_$V3 in the data headers) 
as shown in Equation \ref{eqn_4}. 
Another parameter, $\chi$, containing information 
about the camera geometry which is derived from the 
design specification, has also been corrected. 
For the WFC, $\chi=-38.2^\circ$. 
The degree of polarization of  
ASASSN-15lh at t$_p$=59 days after the $V$-band maximum 
light is shown in Table \ref{table_pol}.
\begin{equation}
\mathrm{p\%=\frac{\sqrt{Q^2+U^2}}{I} \times 
\frac{T_{par}+T_{perp}}{T_{par}-T_{perp}} \times 100 \%}
\label{eqn_3}
\end{equation}
\begin{equation}
\mathrm{P.A.=\frac{1}{2}tan^{-1} \bigg( \frac{U}{Q} \bigg)+PA\_V3+\chi}
\label{eqn_4}
\end{equation}
The classical method proposed by \citet{Serkowski_1958} 
and \citet{Serkowski_1962} is often used for the 
determinations of the polarization and associated uncertainties. 
\citet{Montier_etal_2015} investigated the statistical behavior 
of basic polarization fraction and angle measurements. 
Asymmetrical terms and correlations in the covariance matrix 
have been included (compare to classical determinations 
discussed by \citet{Naghizadeh_Clarke_1993}). 
We use Equation \ref{eqn_5} and Equation \ref{eqn_6} 
to describe the uncertainty of $p\%$ and P.A. 
The detailed derivation has been provided in 
Appendix F of \citet{Montier_etal_2015}. 
\begin{equation}
\mathrm{\sigma_p ^2 \ = \  \frac{1}{p^2 I^4} 
\times \big{(} Q^2 \sigma_Q ^2 + U^2 \sigma_U ^2 + p^4 I^2 \sigma_I ^2 
+ 2QU \sigma_{QU} - 2IQp^2 \sigma_{IQ} -2IUp^2 \sigma_{IU}^2 \big{)}}
\label{eqn_5}
\end{equation}
\begin{equation}
\mathrm{\sigma_{P.A} \ = \  \sqrt{\frac{Q^2\sigma_U ^2 
+ U^2\sigma_Q ^2 -2QU\sigma_{QU}}{Q^2\sigma_Q ^2 
+ U^2\sigma_U ^2 +2QU\sigma_{QU}}} 
\times \frac{\sigma_p}{2p} \  rad}
\label{eqn_6}
\end{equation}
The Stokes I vector gives the total intensity of the source. 
The magnitudes of the SN were obtained by 
applying the ACS/WFC zeropoint correction. 
The aperture corrections calculated with the ACS/WFC 
encircled energy profile for each bandpass has also 
been done according to \citet{Sirianni_etal_2005}. \\
\\

%F435UV:pol60/pol0=0.9741, POL120/POL0=1.0274
%F775W: POL60/POL0=0.9788, POL120/POL0=0.9959
\begin{deluxetable}{lll}
\tablewidth{0pc}
\tabletypesize{\scriptsize}
\tablecaption{Count Rate Ratios used as Correction Factors
\label{table_correction}}
\tablehead{
 \colhead{Band} & \colhead{POL60/POL0} & \colhead{POL120/POL0} \\ \vspace{-0.1cm}  }
\startdata
F435W/POLUV & 1.0266 & 0.9734 \\
F606W/POLV  & 0.979  & 1.014  \\
F775W/POLV  & 0.9788 & 0.9959 \\
\enddata
\end{deluxetable}

%%%%%%%%%%%%%%%%%%%%%%%%%%%%%%%%%%%%%%%%%%%%%%%%%%%%%%%%%%%%%%%%%%%%%%%%%%%%%%%%%

%%%%%%%%%%%%%%%%%%%%%%%%%%%%%%%%%%%%%%%%%%%%%%%%%%%%%%%%%%%%%%%%

\end{document}